\documentclass[floatfix,superscriptaddress,twocolumn,showpacs,prb,amsmath,amssymb]{revtex4}

\usepackage{graphicx}
\usepackage{dcolumn}
\usepackage{bm}
\usepackage{longtable}
\begin{document}

\preprint{APS/123-QED}

\title{Spin dynamics and spin freezing behavior in the two-dimensional antiferromagnet NiGa$_{2}$S$_{4}$ revealed by Ga-NMR, NQR and $\mu$SR measurements}

\author{Hideo Takeya}
\affiliation{Department of Physics, Graduate School of Science, Kyoto University, Kyoto 606-8502, Japan.}

\author{Kenji Ishida}
\email{kishida@scphys.kyoto-u.ac.jp}
\affiliation{Department of Physics, Graduate School of Science, Kyoto University, Kyoto 606-8502, Japan.}

\author{Kentaro Kitagawa}
\affiliation{Department of Physics, Graduate School of Science, Kyoto University, Kyoto 606-8502, Japan.}
\affiliation{Institute for Solid State Physics, University of Tokyo, Kashiwa 277-8581, Japan.}

\author{Yoshihiko Ihara}
\author{Keisuke Onuma}
\author{Yoshiteru Maeno}
\affiliation{Department of Physics, Graduate School of Science, Kyoto University, Kyoto 606-8502, Japan.}

\author{Yusuke Nambu}
\author{Satoru Nakatsuji}%
\affiliation{Department of Physics, Graduate School of Science, Kyoto University, Kyoto 606-8502, Japan.}
\affiliation{Institute for Solid State Physics, University of Tokyo, Kashiwa 277-8581, Japan.}

\author{Douglas E. MacLaughlin}%
\affiliation{Department of Physics, University of California, Riverside, California 92521-0413, USA.}

\author{Akihiko Koda}
\author{Ryosuke Kadono}
\affiliation{Meson Science Laboratory, Institute of Materials Structure Science, High Energy Accelerator Research Organization (KEK), 1-1 Oho, Tsukuba, Ibaraki 305-0801, Japan}

\date{\today}

\begin{abstract}
We have performed $^{69,71}$Ga nuclear magnetic resonance (NMR) and nuclear quadrupole resonance (NQR) and muon spin rotation/resonance on the quasi two-dimensional antiferromagnet (AFM) NiGa$_2$S$_4$, in order to investigate its spin dynamics and magnetic state at low temperatures. 
Although there exists only one crystallographic site for Ga in NiGa$_2$S$_4$, we found two distinct Ga signals by NMR and NQR. The origin of the two Ga signals is not fully understood, but possibly due to stacking faults along the $c$ axis which induce additional broad Ga NMR and NQR signals with different local symmetries.  
We found the novel spin freezing occurring at $T_{\rm f}$, at which the specific heat shows a maximum, from a clear divergent behavior of the nuclear spin-lattice relaxation rate $1/T_{1}$ and nuclear spin-spin relaxation rate $1/T_{2}$ measured by Ga-NQR as well as the muon spin relaxation rate $\lambda$. The main sharp NQR peaks exhibit a stronger tendency of divergence, compared with the weak broader spectral peaks, indicating that the spin freezing is intrinsic in NiGa$_2$S$_4$. The behavior of these relaxation rates strongly suggests that the Ni spin fluctuations slow down towards $T_{\rm f}$, and the temperature range of the divergence is anomalously wider than that in a conventional magnetic ordering. 
A broad structureless spectrum and multi-component $T_1$ were observed below 2 K, indicating that a static magnetic state with incommensurate magnetic correlations or inhomogeneously distributed moments is realized at low temperatures. However, the wide temperature region between 2 K and $T_{\rm f}$, where the NQR signal was not observed, suggests that the Ni spins do not freeze immediately below $T_{\rm f}$, but keep fluctuating down to 2 K with the MHz frequency range. Below 0.5 K, all components of $1/T_1$ follow a $T^3$ behavior. 
We also found that $1/T_1$ and $1/T_2$ show the same temperature dependence above $T_{\rm f}$ but different temperature dependence below 0.8 K. These results suggest that the spin dynamics is isotropic above $T_{\rm f}$, which is characteristic of the Heisenberg spin system, and becomes anisotropic below 0.8 K.       
\end{abstract}

\pacs{Valid PACS appear here}
\maketitle

\section{Introduction}

Recently, vigorous theoretical and experimental studies have been performed on various compounds with geometrically frustrating lattices.\cite{Ramirez} Among such compounds, a two-dimensional (2D) triangular lattice is quite intriguing because it is the simplest and most fundamental structure. Quite recently, a new quasi 2D triangular antiferromagnet NiGa$_2$S$_4$ was discovered as the first example of a bulk low-spin antiferromagnet with an exact triangular lattice.\cite{NakatsujiScience}
NiGa$_2$S$_4$ is a layered compound with the central NiS$_2$ block layers of edge-sharing NiS$_6$ octahedra and top and bottom sheets of GaS$_4$ tetrahedra (Fig.~1). These slabs are stacked along the $c$ axis and connected with each other by a weak van der Waals force. Since the NiS$_2$ layers are effectively decoupled, and the Ni-Ni distance along the $c$ axis is more than three times longer than the Ni-Ni distance along the $a$ axis, NiGa$_2$S$_4$ has been regarded as a nearly ideal 2-D triangular system.  
The electronic configuration of magnetic Ni$^{2+}$ ($3d^8$) ions is $t_{2\rm g}^{6}, e_{\rm g}^{2}$ ($S = 1$). 
Despite strong antiferromagnetic interactions (the Weiss temperature $\theta _{\rm w}\simeq -80$ K), it was reported that no long-range magnetic order is detected down to 0.35 K by susceptibility, specific-heat, and neutron-diffraction measurements.\cite{NakatsujiScience} Instead, incommensurate short-range order with nanoscale correlation, whose wave vector is $\mbox{\boldmath$q$} = (\eta, \eta, 0)$ with $\eta = 0.158 \sim 1/6$, was revealed by the neutron diffraction. In addition, a quadratic temperature dependence of the specific heat below 4 K indicates the presence of coherent gapless linearly dispersive modes at low temperatures, although the bulk measurements show a small anomaly around 10 K (= $T_{\rm f}$) Ref.[\onlinecite{NakatsujiScience}]. However, the origin of the $T^2$ dependence of the specific heat ($C$) has not been understood yet.

In this paper, we report the results of Ga nuclear-magnetic-resonance (NMR) and nuclear-quadrupole-resonance (NQR) measurements on NiGa$_2$S$_4$, which have been performed in order to investigate from a microscopic viewpoint the origin of the anomaly at $T_{\rm f}$ = 10 K, spin dynamics at low temperatures, and the magnetic ground state. We also report the results of muon-spin-rotation/relaxation ($\mu$SR) measurements, which detect the slow dynamics beyond the NMR limitation. We found that the nuclear spin-lattice relaxation rate $1/T_{1}$, nuclear spin-spin relaxation rate $1/T_2$ and the muon spin relaxation rate $\lambda$ all exhibit a clear divergence in approaching from above $T_{\rm f} \sim$ 10 K. In addition, a broad structureless Ga-NQR spectrum and inhomogeneous distribution of $T_1$ were observed below 2 K. These results strongly suggest that the Ni moments freeze out below $T_{\rm f}$ and give rise to inhomogeneous internal fields at the Ga sites below 2 K. Moreover, it was found that the spin dynamics is isotropic above $T_{\rm f}$, which is shown from the identical temperature dependence of $1/T_1$ and $1/T_2$. Isotropic spin dynamics is considered to be characteristics of Heisenberg spin system. The different temperature dependence was observed at low temperatures between $1/T_1$ and $1/T_2$: $1/T_1$ follows a $T^3$ dependence below 0.5 K, and $1/T_2$ is almost linear in $T$ below 2 K. The different temperature dependence suggests anisotropic spin dynamics at low temperatures.

\begin{figure}[htbp]
 \begin{center}
   \includegraphics[width=60mm,clip,angle = 90]{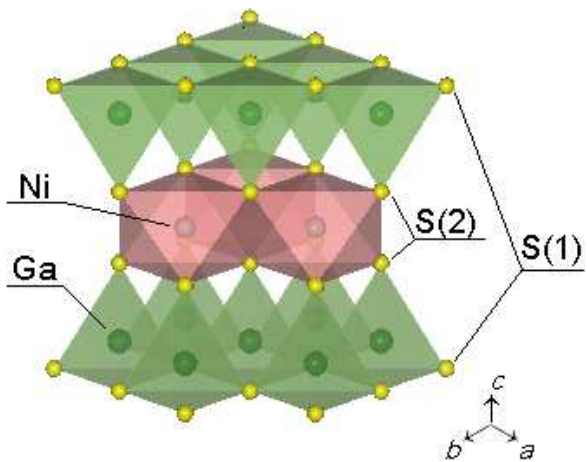}
 \caption{(Color online) Crystal structure of NiGa$_2$S$_4$. The lattice constants are $a = 3.624$ \AA, $c = 11.996$ \AA.}
 \end{center}
\end{figure}

\section{Experimental}
Polycrystalline and single-crystal samples of NiGa$_2$S$_4$ were synthesized as described in the literature.\cite{NakatsujiScience}
Powder x-ray measurements at room temperature and neutron diffraction measurements in the temperature range between 1.5 K and 300 K confirmed that NiGa$_2$S$_4$ retains the trigonal crystal structure down to 1.5 K with $P\bar{3}m1$ symmetry.\cite{NakatsujiScience}
Three samples (polycrystalline and as-grown single crystalline samples of nominal NiGa$_2$S$_4$, and a polycrystalline sample of NiGa$_2$S$_{4.04}$ ) were used in the measurements. The same batch of polycrystalline NiGa$_2$S$_4$ used in the neutron diffraction\cite{NakatsujiScience} was measured in the present NQR experiments, and the single crystal was used in NMR. Since NQR results (NQR spectrum and $1/T_1$) are essentially the same in the three samples, we consider that the NQR results are determined by the intrinsic magnetic properties, which are not affected by a spurious impurity phase.  

The NMR/NQR measurements were performed on two Ga isotopes [$^{69}$Ga ($I = 3/2$): $^{69}\gamma = 10.219$ MHz/10kOe and $^{69}Q = 0.19 \times 10^{-24}$cm$^{2}$, $^{71}$Ga ($I=3/2$) : $^{71}\gamma = 12.984$ MHz/10kOe and $^{71}Q = 0.16 \times 10^{-24}$cm$^{2}$, where $\gamma$ and $Q$ are the gyromagnetic ratio and the electric quadrupole moment, respectively]. NMR/NQR spectra, $1/T_1$ and $1/T_2$ were measured by Ga spin-echo signals in the temperature range of 75 mK - 200 K.   
Zero-field and longitudinal-field $\mu$SR measurements in the temperature range of 1.8 K to 250 K were performed at the $\pi$A-port of the Meson Science Laboratory at KEK in Tsukuba, Japan. The powder sample was attached to a silver``cold plate" by GE-varnish. A $^4$He gas flow cryostat was used for the $\mu$SR measurements. 

\section{Experimental Results}
\subsection{NMR and NQR spectra}
NMR spectra of a single-crystalline NiGa$_2$S$_4$ were obtained by sweeping the external field.
Figure 2 (a) displays the Ga-NMR spectra against the field at 40, 17 and 1.6 K.
Here, the NMR frequency is fixed at 93.5 MHz and the external magnetic field is applied parallel to the $c$-axis.
\begin{figure}[t]
\begin{center}
\includegraphics[clip=,width=0.9\columnwidth]{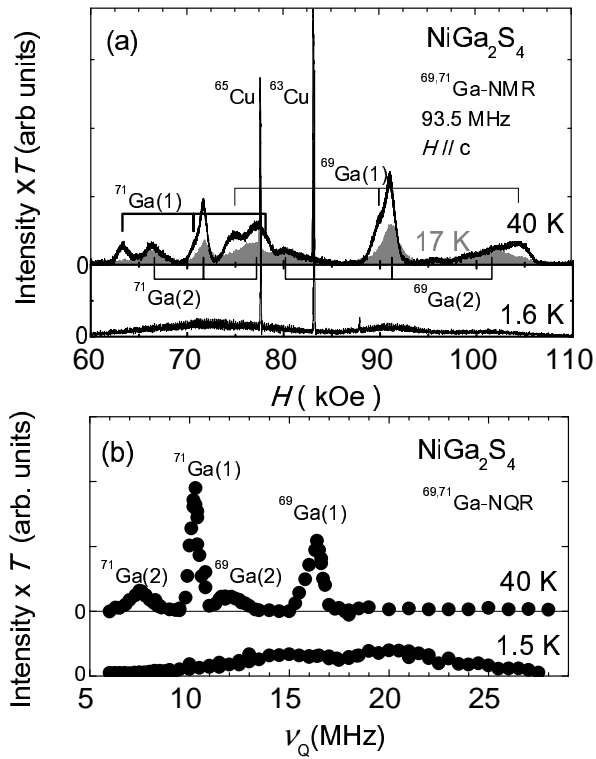}
\caption
{(a)$^{69,71}$Ga-NMR spectra from single-crystal NiGa$_2$S$_4$ at 40 K (curve in upper part), 17 K (gray area), and 1.6 K (lower part). NMR intensity normalization, i.e. taking into account difference in the $T_2$ relaxation rate ($T_2$ correction) has not been made. The resonant frequency is 93.5 MHz and the magnetic field is applied along the $c$-axis.
The time between 1st and 2nd pulses to observe spin echo is 30 $\mu$sec. The $^{63,65}$ Cu signals are from an NMR coil. 
(b)$^{69,71}$Ga-NQR spectra after $T_2$ correction at 40 and 1.5 K using polycrystalline samples. NQR intensity is normalized by temperature. 
}
\end{center}
\end{figure}

In general, NMR spectra for nuclear spin $I (= 3/2)$ show three peaks, which are composed of an intense central peak arising from the $1/2 \leftrightarrow -1/2$ transition and two satellite peaks from $3/2 \leftrightarrow 1/2$ and $-1/2 \leftrightarrow -3/2$ transitions when electric field gradient exists at the observed nucleus site.
Since NiGa$_2$S$_4$ has one crystallographic Ga site surrounded by four S atoms, which forms a GaS$_4$ tetrahedron, three peaks should be observed for each of the $^{69}$Ga and $^{71}$Ga nuclei.
However, the spectrum at 40 K exhibits asymmetric center peaks and two asymmetric satellite peaks, indicative of the existence of two Ga sites (Ga(1) and Ga(2) sites) with different EFGs. With decreasing temperature, the intensity of the NMR signals from the Ga(1) site decreases, and NMR signals from only one Ga site were observed at 17 K as shown by the dark gray area in Fig.~2 (a). It should be noted that measurement of $T_2$ on NMR spectra was needed in order to discuss the intrinsic intensity ratio of the two Ga sites. By further cooling, all NMR signals disappeared below $T_{\rm f} \sim$ 10 K, and extremely broad NMR spectrum reappears below 2 K as shown in the bottom of Fig.~2 (a). The extremely broad Ga-NMR spectrum strongly suggests the occurrence of static magnetism with inhomogeneous internal field well below $T_{\rm f}$.      

The existence of two Ga sites with different EFGs, which was indicated by the Ga NMR, was confirmed by the NQR measurement. Ga-NQR spectra from polycrystalline sample were obtained by measuring the spin-echo intensity as a function of frequency. Unlike the NMR, the NQR spectrum for the $I = 3/2$ nuclei should consist of a single peak of the $\pm 1/2 \leftrightarrow \pm3/2$ levels. We observed two intense peaks with narrow width and two weak peaks with broad width as shown in Fig.~2 (b). The frequency ratio of two intense peaks ( 16.5 MHz / 10 MHz) is nearly the same as that of two weak peaks (12.5 MHz / 7.5 MHz), which is equal to the ratio of the nuclear quadrupole moment ($^{69}Q/^{71}Q$ = 1.589). This indicates that there are two sets of distinct NQR signals, and reveals the existence of two Ga sites with different EFGs in NiGa$_2$S$_4$.
From the separation between two satellite peaks in the NMR spectra at 40 K, we deduced the NQR frequency along the principal axis ($c$ axis) of the EFG at the Ga site ($\nu_z$). We evaluated the asymmetry parameter $\eta = (\nu_y-\nu_x)/\nu_z$ using the relation
\begin{equation*}
\nu _{\rm Q} = \nu _{\rm z}\hspace{0.25em}\sqrt[]{\mathstrut 1+\frac{1}{3}\eta ^2},
\end{equation*}
where $\nu_Q$ is the NQR resonance frequency in Fig.~2 (b).
The values of $\nu_{\rm Q}$, $\nu_{\rm z}$, and $\eta$ are listed in Table I.

\begin{table}[b]
\caption
{The quadrupole parameters for $^{69,71}$Ga(1) and $^{69,71}$Ga(2) derived from the peak position of the NMR and NQR spectra.
The ratio of quadrupole moment of Ga isotopes which corresponds to that of NQR resonance frequency is $^{69}Q / ^{71}Q =  ^{69}\nu_{Q} / ^{71}\nu _{Q} = 1.589$.}
\begin{tabular}{p{2cm} p{2cm} p{2cm} p{2cm} p{2cm}}
\hline
 & $\nu _z$ [MHz] & $\nu _{\rm Q}$ [MHz] & $\eta$\\ \hline \hline
$^{69}$Ga(1) & 15.87 & 16.26 & 0.39 \\
$^{69}$Ga(2) & 10.95 & 11.97 & 0.77 \\
$^{71}$Ga(1) & 10.05 & 10.25 & 0.35 \\
$^{71}$Ga(2) &  6.95 &  7.54 & 0.73 \\
\hline
\end{tabular}
\end{table}

It is noted that two Ga NQR signals are observed, even though there exists only one crystallographic site for Ga in perfect NiGa$_2$S$_4$. 
From careful Inductively Coupled Plasma (ICP) and scanning electron microscope (SEM) measurements,\cite{NakatsujiNambu} it was revealed that S occupation is $\sim 3.96$, suggesting that the S deficiency is at most $\sim 1 \%$, and that the configuration of the triangular structure is rather good. At present, the origin of the two Ga sites is not identified, but we point out possible inclusion of different stacking units, closely related to the structure of NiGa$_2$S$_4$, might give rise to an additional Ga site with a different EFG. A tiny amount of sulfur deficiency and/or disorder is considered to exist in the `outer' sulfur layer, which is shown by S(1) in Fig.~1. It is considered that the sulfur deficiency and/or disorder is more easily introduced in the outer S(1) layer than in the inner S(2) layers which are coupled strongly with Ni$^{2+}$ ions and form the NiS$_2$ block layer. 
The fraction of the two Ga sites estimated from the Ga-NQR intensity at 40 K, which is normalized by $1/T_2$ values, is Ga(1) : Ga(2) = 0.78 : 0.22. If we assume that the Ga(2) NQR is ascribed to the Ga site which is influenced by the sulfur deficiency at the S(1) site, we estimate that the S(1) deficiency is approximately 7 \% from the intensity ratio of two Ga NQR signal because one S(1) deficiency gives an influence to three Ga atoms in the low concentration limit. However, this possibility might be excluded from the ICP result. Instead, we consider that the stacking faults along the $c$ axis, which result in a different stacking unit from the bulk NiGa$_2$S$_4$, might be the origin of the Ga(2) site. In any case, the linewidth of the Ga(2) NQR signal is twice broader than that of the Ga(1) signal, it is reasonably considered that the Ga(1) signal arises from the Ga site with the regular crystal structure, and the Ga(2) signal arises from the Ga site with disorder and/or defects in the structure. This assignment is consistent with the fact that $\eta$ at the Ga(2) ($\sim 0.75$) is larger than $\eta$ at the Ga(1) ($\sim$ 0.37). 

The NQR-signal intensity at both Ga sites decreases with decreasing temperature and the NQR signals disappear around $T_{\rm f}$ $\sim 10$ K, indicative of a magnetic anomaly.
On further cooling, enormously broad NQR spectra were observed as in the NMR spectra below 2 K. Since no obvious structure was found in the spectra, it is inferred that the internal field at the Ga sites is widely distributed in a static magnetic state at low temperatures. Such magnetic state is discussed in section IV B, C.

\subsection{Knight Shift}

Next, we show the temperature dependence of the Knight shift ($K$) measured at the central peaks of the Ga-NMR spectrum. In the $K$ estimation, the shift originating from the 2nd-order quadrupole interaction was subtracted.  Figure 3 displays the temperature dependence of $K(T)$ for $^{71}$Ga(1) and $^{71}$Ga(2), along with the behavior of the bulk susceptibility normalized by the behavior of the Knight shift above 100 K.\cite{NakatsujiScience}  
In general, $K(T)$ is decomposed as,
\begin{equation*}
K(T) = K_{\rm spin}(T) + K_{\rm orb},
\end{equation*}
where $K_{\rm spin}(T)$ and $K_{\rm orb}$ are the spin and orbital parts of Knight shift, respectively.
$K_{\rm spin}$ is related to the bulk susceptibility $\chi(T)$ originating from the Ni spins as 
\begin{equation*}
K_{\rm spin}(T) = \frac{A_{\rm hf}}{N_{\rm A}\mu _{\rm B}}\chi(T),
\end{equation*}
where $A_{\rm hf}$, $N_{\rm A}$, and $\mu_{\rm B}$ are the hyperfine coupling constant between the Ga-nuclear spin and Ni-3$d$ spins, Avogadro's number and the Bohr magneton, respectively.
$K_{\rm orb}$ is related with the Van Vleck susceptibility, which is temperature independent in general.

As shown in Fig.~3, the Knight shift at both sites follows the bulk susceptibility above 80 K. The hyperfine coupling constant $A_{\rm hf}$ at both sites [$17.70\pm 0.20$ (kOe/$\mu_{\rm B}$) for Ga(1), $7.67\pm 0.36$ (kOe/$\mu_{\rm B}$) for Ga(2)] was evaluated from the slopes in the $K$-$\chi$ plot displayed in the inset of Fig.~3. However, $K$ at the Ga(1) site could not be measured below 50 K due to the decrease of the NMR-signal intensity, and $K$ at the Ga(2) site deviates from the temperature dependence of $\chi(T)$ below 80 K. The Knight shift probing the microscopic susceptibility decreases although the bulk susceptibility continues to increase. It seems that a precursor of the magnetic anomaly starts below 80 K. This corresponds well to the Weiss temperature obtained from $\chi(T)$. We point out that a similar deviation of $K(T)$ from $\chi(T)$ at low temperatures was observed in a number of geometrically frustrated spin systems such as SrCr$_8$Ga$_4$O$_{19}$(SCGO),\cite{MendelsPRL00} Ba$_2$Sn$_2$ZnCr$_{7p}$Ga$_{10-7p}$O$_{22}$~\cite{BonoPRL04} and FeSc$_2$S$_4$.\cite{ButtgenPRB06} In these compounds, the possibility of spin singlet state and/or gapped state with the defect like contribution has been suggested. However such possibilities are not applicable to the present case because the divergence of $1/T_1$, which is discussed later, suggests a magnetic ground state. In the magnetic state, there exist static moments giving rise to inhomogeneous internal fields at the Ga sites. The different temperature dependence between microscopic and macroscopic spin susceptibilities in the highly frustrated compounds, which are probed with NMR and bulk susceptibility respectively, remains unclear, and deserves to be understood theoretically.              
\begin{figure}[t]
 \begin{center}
\includegraphics[clip=,width=0.9\columnwidth]{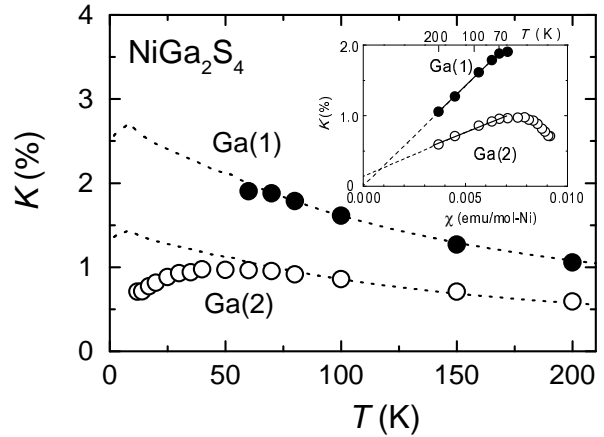}
\caption{Temperature dependence of the Knight Shift $K$ for $^{71}$Ga(1)($\bullet$) and $^{71}$Ga(2)($\circ$) sites.
The dotted curves in the main panel give the temperature dependence of susceptibility $\chi$ normalized to $K$ above 100 K.
Inset: $K$ vs. $\chi$ plot for the two $^{71}$Ga sites. Dotted lines are the linear fits between 70 K and 200 K.
}
 \end{center}
\end{figure}

\subsection{Nuclear spin-lattice and spin-spin relaxation rate, $1/T_1$ and $1/T_2$}

The nuclear spin-lattice relaxation rate (longitudinal rate) $1/T_1$ was measured by the saturation-recovery method with saturation pulses. Above $T_{\rm f}$, the recovery curve $m(t)$ of the nuclear magnetization $M(t)$ at time $t$ after the saturation pulses, which is defined by $m(t)=[M(\infty)-M(t)]/M(\infty)$, is consistently fitted by the $m(t) \propto \exp(-3t/T_1)$ at both Ga sites as shown on a semilogarithmic scale in Fig.~4 (a). Thus a single $T_1$ component was determined above 10 K. 
The nuclear spin-spin relaxation rate (transverse rate) $1/T_2$ was measured by recording the spin-echo intensity $I(2\tau)$ by changing the time separation $\tau$ between the $\pi/2$ (1st) and $\pi$ (2nd) pulses. $I(2\tau)$ was well fitted by the relation of $I(2\tau) \propto \exp(-2\tau/T_2)$ at both Ga sites as shown in Fig.~4 (b). Here, the pulse length of a $\pi/2$ is approximately 5 $\mu$sec.

\begin{figure}[b]
\begin{center}
\includegraphics[clip=,width=0.9\columnwidth]{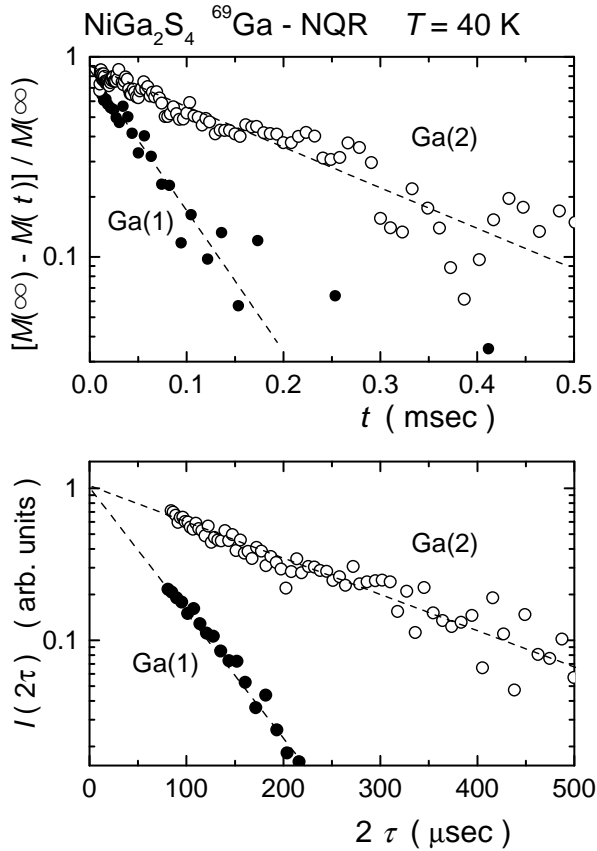}
\caption{(a) Recovery curves $m(t)$ derived from the $^{69}$Ga nuclear magnetization at two Ga sites ($^{69}$Ga(1)($\bullet$) and $^{69}$Ga(2)($\circ$)) by changing the time interval between a saturation pulse and 1st pulse. Single component of $T_1$ was derived at two sites. (b) Decay curve $I(2\tau)$ from the $^{69}$Ga nuclear magnetization obtained by changing the duration time $\tau$ between 1st and 2nd pulses. The decay curves are consistently fitted by an exponential function [$I(2\tau) \propto \exp(-2\tau/T_2)$].}
\end{center}
\end{figure}
\begin{figure}[htbp]
\begin{center}
\includegraphics[clip=,width=0.9\columnwidth]{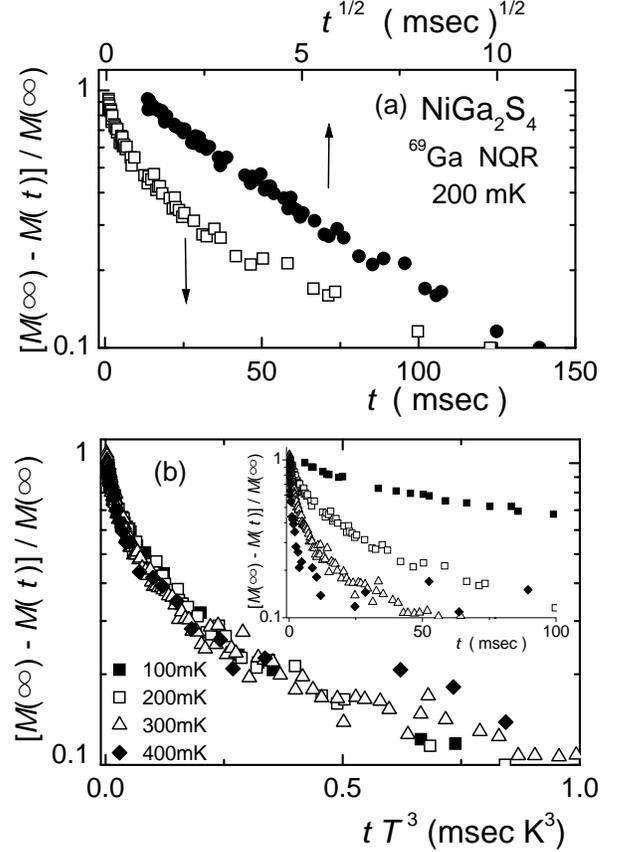}
\caption{(a) Relaxation curves $m(t)$ plotted at 200 mK against $t$ (bottom axis, black circles) and $\sqrt{t}$ (top axis, open circles), where $t$ is the time between the saturation pulse and the 1st spin-echo pulse.
(b) Several recovery curves at several temperatures far below $T_{\rm f}$ plotted against $tT^{ 3}$, where $T$ is the temperature. The inset of Fig.~5 (b) shows the same recovery curves plotted against $t$.}
 \end{center}
\end{figure}
Between $T_{\rm f}$ and 2 K the NQR signals are not observed due to the extremely short $T_1$ and $T_2$. This is due to the critical slowing down of the spin fluctuations to the NQR frequency scale, which results in the rapid spin relaxation. On further cooling, a broad NQR spectra in Fig.~2 (b) was observed below 2 K. The $m(t)$ exhibits the multi-component behavior, which is characterized by the upward curve as shown in Fig.~5 (a), which $m(t)$ is consistently fitted by the relation $m(t) \propto \exp(-\sqrt{3t/\tilde{T_1}})$ in Fig. 5 (a). This is shown by the straight line in the semi-log plot between $m(t)$ and  $\sqrt{t}$. This relation is often observed when $1/T_1$ is inhomogeneously distributed due to the presence of the relaxation center, e.g. a amount of doped magnetic impurities and/or crystal imperfection.\cite{McHenryPRB72} In such a case, $1/\tilde{T_1}$ is normally adopted as $1/T_1$. On the other hand, it was found that $I(2\tau)$ follows the same $I(2\tau) \propto \exp(-2\tau/T_2)$ relation even below $T_{\rm f}$. The inset of Fig.~6 shows the plot of the $I(2\tau)$ against 2$\tau$ at 76, 220, and 485 mK. Thus single-component $1/T_2$ was determined down to the lowest temperature.

\begin{figure}[htbp]
\begin{center}
\includegraphics[clip=,width=0.9\columnwidth]{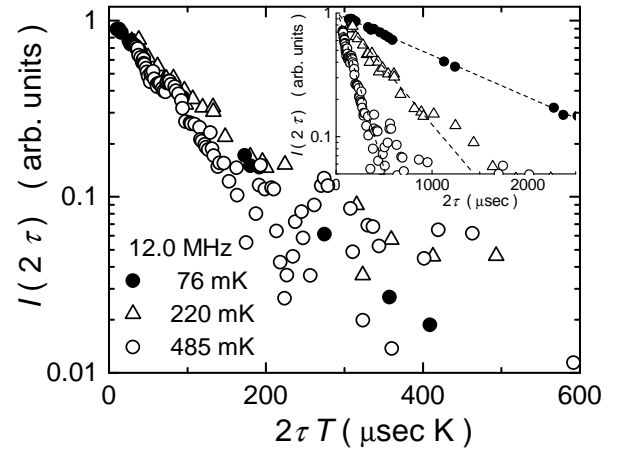}
\caption{Ga-NQR decay curves $I(2\tau)$ of NiGa$_2$S$_4$ are plotted against $2\tau T$ at various temperatures far below $T_{\rm f}$. The inset shows the same decay curves plotted against $t$.}
 \end{center}
\end{figure}

\begin{figure}[t]
\begin{center}
\includegraphics[clip=,width=0.9\columnwidth]{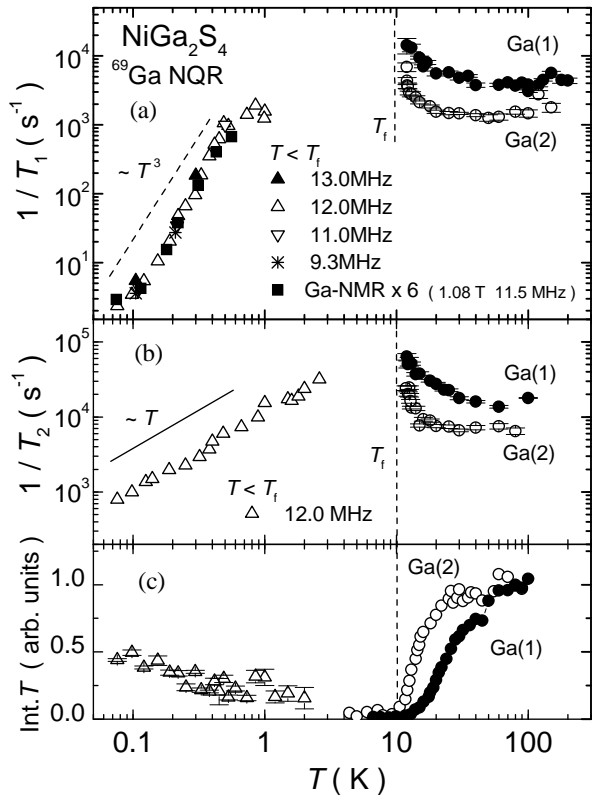}
\caption{(a) Temperature dependence of nuclear spin-lattice relaxation rate 1/$T_1$. 1/$T_1$ was measured at $^{69}$Ga(1)($\bullet$), $^{69}$Ga(2)($\circ$) site above $T_{\rm f}$. Below 1 K, $1/\tilde{T_1}$ was derived from the fitting of $m(t)$ measured at four frequencies in a range between 9.5 MHz and 13 MHz. Solid squares give 1/$\tilde{T_1}$ measured at 1.08 T with resonant frequency of 11.5 MHz. $1/T_1$ for NMR is normalized for the difference in the matrix elements (see text). (b) Temperature dependence of nuclear spin-spin relaxation rate $1/T_2$. Below 2.5 K, $1/T_2$ was measured at 12 MHz in the broad NQR spectrum affected by the inhomogeneous internal fields. Decay curves below 2.5 K are consistently fitted by a single exponential function down to a lowest temperature as shown in Fig.~6 although $1/T_1$ is widely distributed as shown in Fig. 5 (b). (c) Temperature dependence of integrated intensities of spin echo after $T_2$ correction. Below 2.5 K, we assume that $T_2$ is independent of frequency. }
 \end{center}
\end{figure}

The temperature dependences of $1/T_1$, $1/\tilde{T_1}$ and $1/T_2$ determined by above fitting procedures are displayed in Figs. 7 (a) and 7 (b), together with the temperature dependence of the integrated NQR intensity multiplied by temperature, which is normalized to the value at 100 K [Fig.~7 (c)]. A $T_2$ correction has been made for the estimation of the NQR intensity in Fig.~7 (c), and we assume that $T_2$ is independent of frequency below 2.5 K.                  
Figures 7 (a) and (b) display the temperature dependence of 1/$T_1$ and $1/T_2$ measured at the two $^{69}$Ga sites above 10 K, $1/\tilde{T_1}$ measured at various frequencies, and $1/T_2$ at 12 MHz, respectively. 
The ratio of $1/T_1$ between the two isotopes at high temperatures is ($1/^{69}T_1$)/($1/^{71}T_1$) $=$ 0.67 $\pm$ 0.05, which is in good agreement with the ratio of gyromagnetic ratio ($^{69}\gamma$/$^{71}\gamma$)$^2$ = 0.62. This indicates that $1/T_1$ is dominated by the magnetic interaction with the electrons, not by the electric quadrupole interaction.
Above 80 K, $1/T_1$ of $^{69}$Ga(1) is nearly constant, which is often observed in a localized-moment system when the temperature is higher than the exchange energy between the localized moments.
Below 80 K, as shown in Fig.~7 (a), $1/T_1$ gradually increases with decreasing temperature, and correspondingly the intensity of the NQR signal decreases. Such  behavior is considered as a precursor of the spin freezing of Ni-$3d$ spins.
If we compare the temperature dependence of $1/T_1$ at the two Ga sites, $^{69}$Ga(1) and $^{69}$Ga(2) sites, the magnitude of $1/T_1$ at the Ga(1) site is three times larger than $1/T_1$ at the Ga(2) site. It should also be noted that the onset temperature of the divergent behavior below about 80 K is higher at the Ga(1) site than at the Ga(2) site. From the comparison of the temperature dependence of $1/T_1$ at the two Ga sites, we conclude that the magnetic anomaly seen in $T_{\rm f}$ is an intrinsic nature of NiGa$_2$S$_4$, because the Ga(1) site possesses a narrower NQR spectrum, indicative of the homogeneous site. If $T_{\rm f}$ were induced by inhomogeneity of the sample, such as sulfur disorder and/or stacking faults, $1/T_1$ at the Ga(2) site, where larger inhomogeneity is suggested by the NQR spectrum, would show the magnetic precursor from higher temperatures. Obviously, this is not the case. A similar difference between the Ga(1) and Ga(2) sites, which indicates that the magnetic anomaly at $T_{\rm f}$ is intrinsic in NiGa$_2$S$_4$, was also observed in $1/T_2$ and NQR-intensity results as shown in Figs.~7 (b) and (c).

\begin{figure}[htbp]
\begin{center}
\includegraphics[clip=,width=0.9\columnwidth]{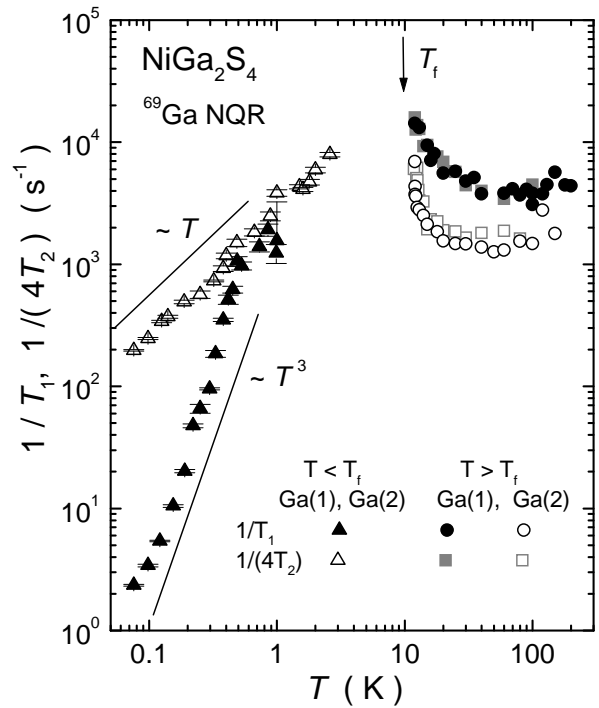}
 \caption{Temperature dependence of the NQR relaxation rate $1/T_1$ and $1/(4T_2)$ at $^{69}$Ga(1) and $^{69}$Ga(2) sites. }
 \end{center}
\end{figure}

As noted above, in the temperature range between 10 and 2 K, the NQR signal was not observed due to the extremely short $T_1$ and $T_2$ beyond the limit of the NMR measurement. Below 2 K, the relaxation rate is distributed inhomogeneously in the low-temperature region. This multi-component $T_1$ behavior is often observed in the spin-glass\cite{CurroPRL00} and Kondo-disordered systems.\cite{MacLaughlinPRL01} $1/\tilde{T_1}$ determined from $m(t) \propto \exp\left(-\sqrt{3t/\tilde{T_1}}\right)$ decreases strongly with decreasing temperature as shown in Fig.~7 (a). Here, $\tilde{T_1}/3$ is the time when $m(t)$ decays to $1/e$, is regarded as a characteristic value of $T_1$ in multi-component $T_1$. $1/\tilde{T_1}$ was measured at several NQR frequencies from 9.5 and 28 MHz at 1.5 K, but the values of $\tilde{T_1}$ do not change so much in this frequency range. The temperature dependence of $1/\tilde{T_1}$ below 2 K is shown in Fig.~7 (a) in the frequency range between 9.3 and 13 MHz, and is independent of the frequency. $1/\tilde{T_1}$ measured in 10 kOe shows the same temperature dependence as that in zero field, although the value of $1/\tilde{T_1}$ in 10 kOe is approximately six times smaller than $1/\tilde{T_1}$ in zero field. This is due to the difference of the matrix elements of the transition probability. 

We found that $1/\tilde{T_1}$ is proportional to $T^{3}$ between 0.8 and 0.1 K. It should be noted that, although the spin dynamics is highly inhomogeneous, the $T^{3}$ dependence holds in all $1/T_1$ components, because the experimental data of $m(t)$ at several temperatures below 0.5 K lie on the same curve when plotted against $tT^3$ as shown in Fig.5 (b). Therefore, the $T^3$ dependence of $1/\tilde{T_1}$ is intrinsic behavior at low temperatures in NiGa$_2$S$_4$.
Temperature dependence of $1/\tilde{T_1}$ is discussed in Sec. IV.

Figure 7 (b) shows temperature dependence of $1/T_2$. $1/T_2$ at both Ga sites also diverges at $T_{\rm f}$, but $1/T_2$ at the Ga(1) site has larger values and starts to diverge at a higher temperature, which is in agreement with the $1/T_1$ results. Below 2.5 K, $1/T_2$ follows a $T$-linear temperature dependence down to the lowest temperature as shown from the plot of $I(2\tau)$ against $2\tau T$ at several temperatures in the main panel of Fig.~6. 
The large $1/T_2$ values and the strong temperature dependence of $1/T_2$ that follows $1/T_1$ behavior above $T_{\rm f}$ indicate that $1/T_2$ is not determined by the nuclear dipole interaction, but by the $T_1$ process. We found that the ratio of $T_2/T_1$ in NiGa$_2$S$_4$ is $4.2 \pm 0.2$ above $T_{\rm f}$. We show temperature dependence of $1/T_1$ and $1/T_2$ divided by 4 in Fig. 8. It should be noted that $1/(4T_2)$ follows quantitatively the same temperature dependence as $1/T_1$ above $T_{\rm f}$ for both Ga sites, and that different temperature dependence is observed in $1/\tilde{T_1}$ and $1/(4T_2)$ below 0.8 K. In general, $1/T_1$ ($1/T_2$) measured with the NQR spectra are affected by the spin dynamics perpendicular (parallel) to the principal axis of EFG. Therefore, it is considered that spin dynamics is isotropic above $T_{\rm f}$, but becomes anisotropic below 0.8 K. We suggest that the isotropic spin dynamics, characterized by the Heisenberg spin system, are slowing down and changed to the anisotropic ones below 0.8 K, where the in-plane spin dynamics is suppressed by the static magnetism, but the out-of-plane spin dynamics remains active and is proportional to temperature. Detailed temperature dependences above and below $T_{\rm f}$ are discussed in Sec.~IV.            

\subsection{Muon spin rotation/relaxation}

Muon spin rotation/relaxation ($\mu$SR) experiments are particularly suitable for detecting {\it slow} spin dynamics with an extremely high relaxation rates beyond the NMR experimental limitation. Therefore, NMR and $\mu$SR are complementary to each other. Detailed $\mu$SR measurements in the temperature range between 2 K and $T_{\rm f}$ published in the literature.\cite{Doug} Here we concentrate on the zero -field $\mu$SR (ZF-$\mu$SR) experiments carried out at the Meson Science Laboratory KEK. 

\begin{figure}[htbp]
\begin{center}
\includegraphics[clip=,width=0.9\columnwidth]{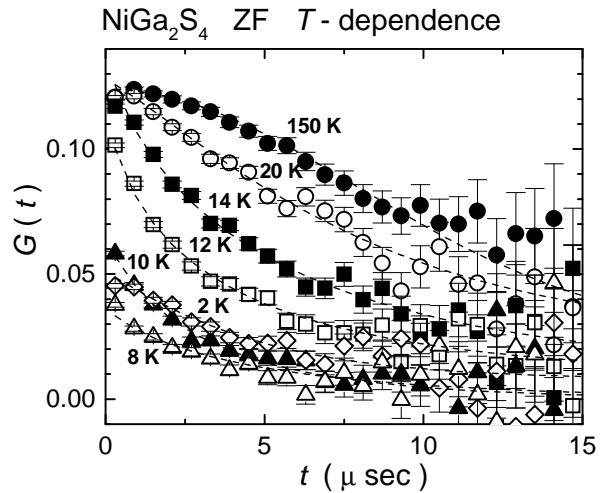}
\caption{Zero-field $\mu$SR asymmetry relaxation functions $G(t)$ in NiGa$_2$S$_4$ at various temperatures. The dotted curves are fits to the power exponential form $G(t) = A \exp{[-(\lambda t)^{\beta}]}$}
 \end{center}
\end{figure}
Figure 9 shows the ZF $\mu$SR asymmetry relaxation function $G(t)$ in NiGa$_2$S$_4$ at various temperatures. The relaxation function is gradually changing from Gaussian to Lorentzian by approaching $T_{\rm f}$. The experimental asymmetry data were fitted to the ``power exponential'' form
\[
G(t) = A \exp{[-(\lambda~t)^\beta]},
\]
where $A$ is the initial muon decay asymmetry, $\lambda$ is a generalized relaxation rate, and the exponent $\beta$ interpolates between exponential ($\beta = 1)$ and Gaussian ($\beta = 2$) limits. This form is suitable for characterizing spin dynamics, i.e. Gaussian (exponential) function suggests that muons see static (dynamics) magnetic fields during their lifetimes ($\sim 2.2 \mu s$). $\beta > 1$ indicates an intermediate dynamics between two limits, and $\beta < 1$ suggests inhomogeneous dynamics with multi-components.

\begin{figure}[htbp]
\begin{center}
\includegraphics[clip=,width=0.9\columnwidth]{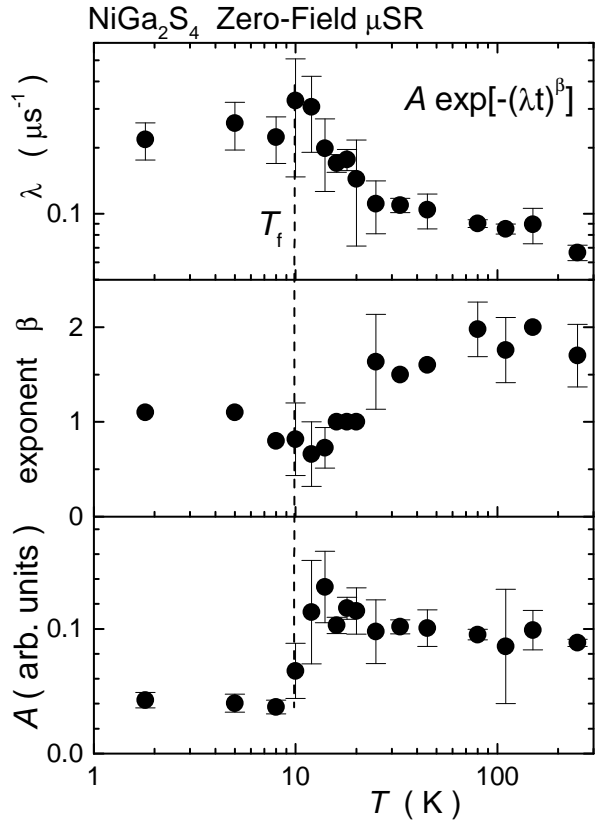}
\caption{Temperature dependence of (a) $\mu$SR relaxation rate ($\lambda$), (b) exponent $\beta$ and (c) asymmetry $A$ derived from the fitting of zero-field asymmetry by $G(t) = A \exp{[-(\lambda~t)^\beta]}$. }
\end{center}
\end{figure}

The relaxation rate $\lambda$, exponent $\beta$, and asymmetry $A$ are plotted against temperature in Fig.~10. Above 100 K, $\lambda$ is small and $\beta$ is close to 2. Both indicate that the nuclear dipole effect is dominant in the relaxation. With decreasing temperature, $\beta$ gradually decreases from 2 to 1, and $\lambda$ increases, although $A$ is nearly constant down to $T_{\rm f}$. These results indicate that $\mu$SR can detect the slow spin dynamics which is beyond the NMR limitation, since the NMR intensity gradually decreases in this temperature range. $\lambda$ shows a distinct peak at $T_{\rm f} \sim 10$ K, below which the asymmetry $A$ decreases to approximately 1/3 of the values above $T_{\rm f}$. As evident from Fig.~9, the so-called ``1/3 tail" is lifted up at 2 K, which indicates that the static field along the implanted muon direction appears at least 2 K. 
In general, the muon spin polarization function below $T_{\rm f}$ is characterized by rapid depolarization of 2/3 of the initial polarization, followed by slow dynamic relaxation of the remaining 1/3 component. This is a characteristic signature of a highly disordered magnetic state in which the moments are quasi-static on the time scale of the muon lifetime.\cite{KT} In the present measurement, the rapid depolarization could not be detected due to the high inhomogeneity of the static moments, but the 1/3 component was detectable. It is considered from the residual asymmetry below $T_{\rm f}$ that almost the entire region in the sample is in the magnetic state. 

It should be noted that relaxation of the 1/3 tail was observed at $T$ = 2 K. Therefore, strong relaxation demonstrates the existence of slow dynamics even below $T_{\rm f}$, which is consistent with the experimental fact that the Ga NQR spectrum is not observed down to 2 K. The present $\mu$SR measurements suggest that magnetic fluctuations of Ni spins are gradually slowing down below 80 K, and that the inhomogeneous static field appears at the implanted muon site at low temperatures. However, fluctuations of the Ni spins remain strong even far below $T_{\rm f}$.
Spin dynamics is discussed in the next section on the basis of NMR and $\mu$SR relaxation behavior.

\section{Discussion}
\subsection{Spin dynamics above $T_{\rm f}$}
In this section, we discuss spin dynamics in NiGa$_2$S$_4$ in the paramagnetic state, which are revealed by $1/T_1$ and $1/T_2$ at the Ga(1) site, because the Ga(1) is considered to be more homogeneous than the Ga(2) site from the narrower NQR spectrum shown in Fig.~2 (b).
Above $T_{\rm f}$, $1/T_1$ and $1/(4T_2)$ behave quantitatively the same as shown in Fig.~8 and 11, and $\lambda$ derived from the $\mu$SR experiments also follows the same temperature dependence.  
\begin{figure}[htbp]
 \begin{center}
 \includegraphics[clip=,width=0.9\columnwidth]{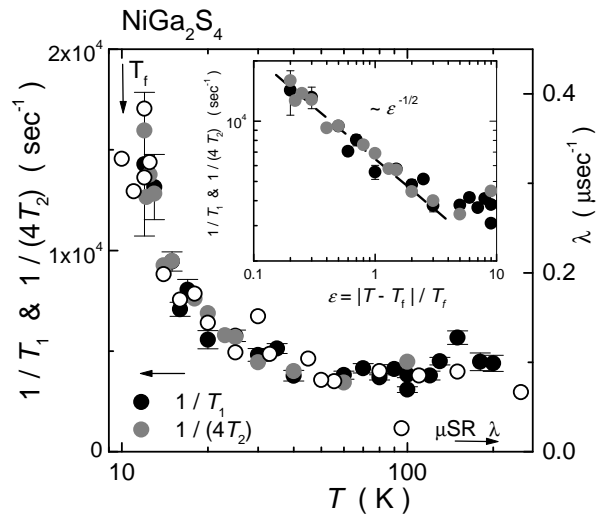}
 \caption{Temperature dependence of $1/T_1$ and $1/(4T_2)$ of the Ga(1) NQR spectra, and $\mu$SR relaxation rate $\lambda$. The inset shows the plot of $1/T_1$ and $1/(4T_2)$ against $(T-T_{\rm f})/T_{\rm f}$.}
 \end{center}
\end{figure}
At temperatures higher than 80 K, $1/T_1$ and $1/(4T_2)$ are nearly constant due to local-moment fluctuations by the exchange interaction between the Ni spins.
Below 80 K, $1/T_1$ and $1/(4T_2)$ start to diverge towards $T_{\rm f}$.  
As seen in the inset of Fig.~11, the divergent behavior of $1/T_1$ and $1/(4T_2)$ is approximately fitted by
\begin{equation*}
1/T_{1,2} \propto \varepsilon^{-\nu}
\end{equation*}
with $\varepsilon=(T-T_{\rm f})/T_{\rm f}$, and $T_{\rm f}$ and $\nu$ are estimated to be $T_{\rm f} = 10.3 \pm 1.0$, and $\nu = 0.45 \pm 0.10$.
According to theoretical prediction, $\nu$ is 0.3 and 0.8 in 3-D and 2-D Heisenberg systems, respectively.\cite{Benner} $\nu \sim 0.5$ suggests that magnetic anomaly might occur due to weak interactions along the $c$ axis, since magnetic order is not realized at a finite temperature in the 2-D Heisenberg system. It should be noted that the critical divergence of the relaxation rate was observed over an unusually wide temperature range $0.2 < \varepsilon < 3$, in which the 2-D short-range correlation is highly developed. This is in marked contrast with the ordinary critical behavior in the 2-D Heisenberg system\cite{Benner}, for which the critical divergence is limited to the vicinity of $T_N$ ($\varepsilon < 0.5$). Such a remarkable divergent behavior persisting over a wide temperature range seems to be one of the key features of the 2-D triangular lattice, because similar critical behavior over wide temperature ranges was reported in LiCrO$_2$ and HCrO$_2$ with 2-D triangular lattices.\cite{AjiroJPSJ88,AlexanderCondmat} 

In addition, it was found that $1/(4T_2)$ shows quantitatively the same temperature dependence as $1/T_1$. This indicates the spin dynamics is isotropic.
When the nuclear dipole interaction is negligibly small, there are the relation between $1/T_1$ and $1/T_2$ measured with the NQR signal arising from the $\pm 1/2 \leftrightarrow \pm 3/2$ of $I$ = 3/2 transitions:\cite{WalstedtPRL67,AulerJPhys96}
\begin{eqnarray*}
\frac{1}{T_1}&=&G_{\perp}(\omega),\\
\frac{1}{T_2}&=&\frac{1}{2}G_{z}(0)+\frac{5}{2}G_{\perp}(\omega).
\end{eqnarray*}
Here, $G_{\alpha}(\omega)$ is the spectral density  of the longitudinal ($\alpha=z$) and transverse ($\alpha = \perp$) components of the fluctuating local field\cite{Slichter}, and is given by:
\begin{equation*}
G_{\alpha}(\omega)=\frac{\gamma^2}{2}\int_{-\infty}^{\infty}\langle\delta H_{\alpha}(t)\delta H_{\alpha}(0)\rangle\exp{(i\omega t)}dt,
\end{equation*}
where $\gamma$ is the gyromagnetic ratio, and $\delta H_z(t)$ ($\delta H_{\perp}(t)$) is the longitudinal (transverse) component of the fluctuating local field. 
If we assume an exponential correlation function,
\begin{equation*}
\langle\delta H_{\alpha}(t)\delta H_{\alpha}(0)\rangle = \langle(\delta H_{\alpha})^2\rangle\exp{\left(\frac{-|t|}{\tau_c}\right)}, 
\end{equation*}
using the correlation time $\tau_c$ of the fluctuation, $1/T_1$ and $1/T_2$ are expressed as follows, 
\begin{eqnarray*}
\frac{1}{T_1}&\propto&\gamma^2\langle(\delta H_{\perp})^2\rangle\frac{\tau_c}{1+(\omega\tau_c)^2}\\
\frac{1}{T_2}&\propto&\frac{1}{2}\gamma^2\langle(\delta H_{z})^2\rangle\tau_c+\frac{5}{2}\gamma^2\langle(\delta H_{\perp})^2\rangle\frac{\tau_c}{1+(\omega\tau_c)^2}.
\end{eqnarray*}
As seen in the equations, $1/T_1$ and $1/T_2$ are determined by the dimensionless value of $\omega\tau_c$, and are proportional to $\tau_c$ (1/$\omega^2\tau_c$) in the case of $\omega\tau_c \ll 1$ ($\omega\tau_c \gg 1$). When $\tau_c$ becomes longer due to the slowing down of the Ni spins with approaching $T_{\rm f}$, the relaxation rates once increase with a maximum at $\omega\tau_c = 1$ and then decrease with decreasing temperature.       
The experimental facts that $1/T_2 \sim 4(1/T_1)$ and both show quantitatively the same temperature dependence each other imply that the magnetic fluctuations are in the regime $\omega\tau_c \ll 1$ and isotropic, and that the relation between $\delta H_{\perp}$ and $\delta H_{z}$ is $\langle(\delta H_{z})^2\rangle \sim 1.5 \langle(\delta H_{\perp})^2\rangle$ in the temperature region above $T_{\rm f}$. The relaxations behaviors are consistent with the Heisenberg spin system.  

With decreasing temperature, magnetic fluctuations are slowing down below 80 K, and seem to be static below $T_{\rm f}$. Now, we discuss the temperature dependence of the correlation time $\tau_c(T)$ of the magnetic fluctuations quantitatively using the experimental data for $1/T_1$ and $A_{\rm hf}$ at the Ga(1) site and magnetic susceptibility $\chi$. 
In the case when the nuclear spin-lattice relaxation rate at the Ga site is dominated by magnetic fluctuations of the Ni localized spins, $1/T_1$ is expressed as follows;
\begin{equation*}
\frac{1}{T_1} = z \frac{\gamma_n^2~k_B~T}{2\mu_B^2}\lim_{\omega \rightarrow 0} \sum_{q} [A(q)]^2 \frac{\chi''(q,\omega)}{\omega},
\end{equation*}
where $z(=3)$ is the number of the nearest neighbor Ni sites to a Ga site, $A(q)$ is the $q$-dependent hyperfine coupling constant at a Ga site from a Ni spin, and $\chi''(q,\omega)$ is the dynamical susceptibility and the sum is over the Brillouin zone.
At temperatures much higher than $T_{\rm f}$, the spin dynamics is determined by independent Ni moments, and the local-moment ($q$ = 0) susceptibility is given by
\begin{equation*}
\chi_L(\omega) =\frac{\chi_0(T)}{1- i \omega \tau_c(T)},
\end{equation*}
where $\chi_0$ is the magnetic susceptibility per Ni atom (emu / Ni-atom).

We consider that $A(q)$ and the dynamical susceptibility are isotropic because $1/T_1$ and $1/(4T_2)$ show nearly the same behavior in this temperature region, and take $A(q) \sim A_{\rm hf}/z = 17.7/3 = 5.9$ (kOe/$\mu_B$). 
Then $1/T_1$ is described in the regime of $\omega\tau_c \ll 1$ as,
\begin{equation*}
\frac{1}{T_1} = 3\frac{\gamma_n^2k_{\rm B}T}{\mu_{\rm B}^2}A_{\rm hf}^2\chi_0(T)\tau_c(T).
\end{equation*}
Therefore, the characteristic energy of the spin fluctuations $\Gamma/k_{\rm B}$, which corresponds to $\hbar/[\tau_c(T) k_{\rm B}$], is given by
\begin{eqnarray*}
\Gamma/k_{\rm B}=\hbar/[\tau_c(T)k_B] &=& 3\frac{(\gamma_nA_{\rm hf})^2\hbar}{\mu_B^2}\chi_0(T)T_1T.\\
              &=& 1.4 \times 10^5~T_1T \chi_0 N_{\rm A} \hspace*{0.5cm} ( \rm K).
\end{eqnarray*}
\begin{figure}[htbp]
 \begin{center}
 \includegraphics[clip=,width=0.9\columnwidth]{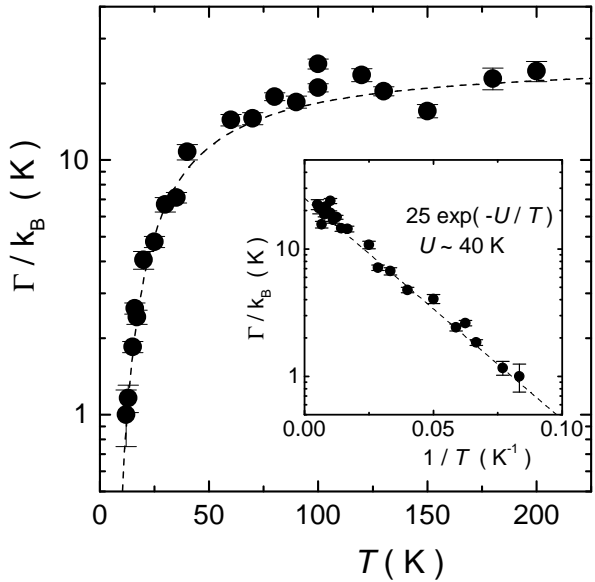}
 \caption{Temperature dependence of the characteristic energy of the spin fluctuations $\Gamma(T)/k_{\rm B}$ (see in text). The inset shows the plot of $\Gamma(T)/k_{\rm B}$ against $1/T$ }
 \end{center}
\end{figure}

Figure 12 shows the temperature dependence of $\Gamma(T)/k_{\rm B} = \hbar/[\tau_c(T)k_{\rm B}$]. $\Gamma(T)/k_B$ is constant above 100 K with the value of $\Gamma(T)/k_{\rm B} \sim 20$ K. This is in good agreement with the exchange interaction $J$ between Ni spins, which is estimated to be $J \sim 20$ K from the relation
\[
J=\frac{3k_{\rm B}\theta_W}{z'S(S+1)},
\]
where we use the Weiss temperature $\theta_W$ of $- 80$ K, and the number of the nearest neighbor Ni sites for a Ni ion $z'$ of 6.

It was found that $\Gamma(T)/k_{\rm B}$ shows thermally activated decrease below 100 K, described by the relation of 
\[
\Gamma(T)/k_{\rm B} = A \exp{(-U/T)}.
\]
From the inset of Fig.~12, $A$ and $U$ are estimated to be 25 K and 40 K, respectively. Here, $U$ corresponds to a binding energy of the local spin configuration determined by $J$.
In general, we expect from the temperature dependence of $1/T_1$ that magnetic correlation length $\xi(T)$ develops exponentially with decreasing temperature, because $\Gamma(T)$ is related to $\xi(T)$ as $\Gamma(T) \propto [c/\xi(T)]^n$ in the Heisenberg spin system. However, neutron experiments showed that the $\xi(T)$ remains a couple of lattice constants even at 1.5 K.\cite{NakatsujiScience} It seems that $\Gamma(T)$ is not directly related to $\xi(T)$ in NiGa$_2$S$_4$. 
Alternatively, we point out that the similar thermally-activated behavior was reported in ESR studies on 2-D triangular compounds of LiCrO$_2$ and HCrO$_2$,\cite{AjiroJPSJ88} and in NMR studies on a Heisenberg kagome lattice antiferromagnet KFe$_3$(OH)$_6$(SO$_4$)$_2$.\cite{NishiyamaPRB03} Particularly, Ajiro {\it et al.}\cite{AjiroJPSJ88} pointed out that the activation energy $U$ in LiCrO$_2$ and HCrO$_2$ takes a universal value equivalent to $U = (2.7 \pm 0.3)JS^2 = 4 k_BT_{\rm KM}$, where $T_{\rm KM}$ is a critical temperature for a Kosterlitz-Thouless-type phase transition. In the 2-D triangular spin system, the possibility of the ``$Z_2$ vortex state'' has been suggested.\cite{KawamuraJPSJ84} Vortices related to the topologically stable defects appear at higher temperatures and starts to coupled each other below $T_{\rm KM}$.
If we apply the above analyses to NiGa$_2$S$_4$, $U \sim 40$ K corresponds to $U \sim  2 JS^2 = 4k_{\rm B}T_{\rm f}$. This relation is quite different with the that observed in dilute-alloy spin glasses.\cite{UemuraPRB85} In these compounds, the correlation time $\tau_c$ obeys the Arrhenius law $\tau_c=\tau_0 \exp(E_a/k_{\rm B}T)$ with a very large activation energy $E_a \sim 20 k_{\rm B}T_{\rm g}$. This indicates that the freezing behavior is observed near $T_g$ and the spins slow down immediately. We point out that the freezing behavior observed in NiGa$_2$S$_4$ is different from the canonical spin-glass systems, but the possibility of the ``$Z_2$ vortex'' transition occurring in NiGa$_2$S$_4$. It is feasible that $T_{\rm f}$ is regarded as $T_{\rm KM}$ in NiGa$_2$S$_4$.

We point out that the spin dynamics in NiGa$_2$S$_4$ possesses the novel 2-D spin character, discussed in the 2-D Heisenberg compounds for nearly two decades.\cite{AjiroJPSJ88} As discussed above, it was revealed that some kind of magnetic correlations start to develop below $|\theta_{\rm W}| \sim 80$ K, which is far above $T_{\rm f}$, and continue to grow down to $T_{\rm f}$. The development of magnetic correlations over wide temperature range and the occurrence of a magnetic anomaly well below $\theta_{\rm W}$ seems to be characteristic of frustrated magnetism, but it is still unclear what kind of magnetic correlations are developing below $\theta_{\rm W}$. The magnetic correlations developing in the frustrated systems is an interesting issue to be studied from experimental and theoretical points of view.

\subsection{Ga-NQR Spectrum below 2 K}  
As shown in Fig.~2 (b), a broad and structureless spectrum was observed at 1.5 K. This spectrum indicates the presence of inhomogeneous static magnetic fields at the Ga sites.
When the internal magnetic field appears at an Ga nuclear sites below $T_{\rm f}$ with the electric quadrupole interaction, the Zeeman interaction from the internal fields is added to the total nuclear Hamiltonian.
The Zeeman interaction is expressed as
\[
{\cal H}_{\rm Z} = -\gamma_{\rm n}\hbar \mbox{\boldmath$I$}\cdot\mbox{\boldmath$H$}_{\rm int},
\]
where $\gamma_{\rm n}$ is the Ga nuclear gyromagnetic ratio and $\mbox{\boldmath$H$}$$_{\rm int}$ is the internal field at the Ga nuclear site. 
The structureless broad spectrum is approximately reproduced by using the inhomogeneous internal field at the Ga nuclear site, as shown in the inset of Fig.~13. The average and distribution width of the internal field at the Ga site are approximately 0.5 T and 0.5 T, respectively, and the calculated Ga-NQR spectrum is shown by the dotted line in the bottom figure of Fig.~13. In the calculation, we assume that the inhomogeneous internal field has only a $c$-axis component for simplicity.
$\mbox{\boldmath$H$}$$_{\rm int}$ arises mainly from the transferred hyperfine field and the dipolar field from the Ni ions, which is expressed as
\[
\mbox{\boldmath$H$}_{\rm int}=\sum_i{\frac{A_{\rm hf}}{N_{\rm A}\mu_B}\mbox{\boldmath$m$}_i}-\sum_i{\frac{1}{\mbox{\boldmath$r$}_i^3}\left\{\mbox{\boldmath$m$}_i+\frac{3\mbox{\boldmath$r$}_i(\mbox{\boldmath$r$}_i\cdot\mbox{\boldmath$m$}_i)}{\mbox{\boldmath$r$}_i^2}\right\}},
\] 
where $m_i$ is the magnetic moment at the $i$-th Ni site and $r_i$ is the vector connecting the Ga site to the $i$-th Ni site. The average value of the ordered moments is estimated to be $\sim 1.0 \mu_B$ from the first term of the right-hand side in the equation above. It is difficult to estimate accurately the ordered moments and the magnetic structure from such a broad spectrum, since the magnitude of the dipolar fields highly depends strongly on the orientations of the moments. 

\begin{figure}[htbp]
 \begin{center}
 \includegraphics[clip=,width=0.9\columnwidth]{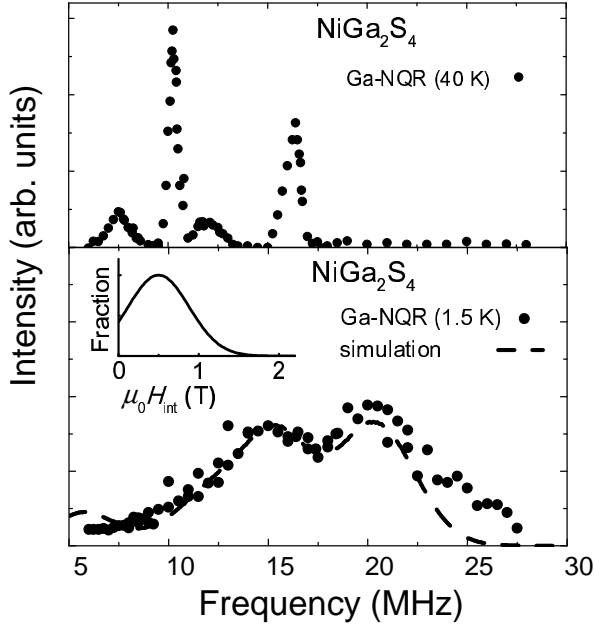}
 \caption{$^{69,71}$Ga-NQR spectra at 40 K (upper panel) and at 1.5 K (bottom panel). The broad spectrum in the bottom figure is approximately reproduced by the presence of inhomogeneous internal fields pointing along the $c$ axis. The distribution of the internal fields is shown in the inset of the bottom figure.    }
 \end{center}
\end{figure}

\subsection{Spin dynamics below $T_{\rm f}$}
Since an extremely broad signal was observed below 2 K, it is concluded that the static magnetic fields are present below 2 K. However, over the wide temperature range below $T_{\rm f}$, slow spin dynamics remains as probed by NQR and $\mu$SR.   
Taking into account the multi-exponential behavior in the relaxation curves at low temperatures, it is concluded that a 2-D inhomogeneous magnetic state is realized in this compound.

We found the spin dynamics becomes anisotropic below 0.8 K from the different temperature dependence of $1/T_1$ and $1/(4T_2)$ as shown in Fig.~8.  
$1/T_1$ follows a $T^3$ dependence in the temperature range between 0.5 and 0.1 K, and this $T^3$ dependence holds for all the spin dynamics detected by the Ga NQR as discussed above.
On the other hand, $1/(4T_2)$ shows a $T$-linear dependence below 1 K. If we take into account that $1/T_1$ and $1/(4T_2)$ probe spin dynamics along different directions, we can conclude that $G_{\perp}(\omega)$ and $G_z(0)$ show the $T^3$ and $T$-linear dependence, respectively. 
It is noteworthy that isotropic spin dynamics above $T_{\rm f}$ change to anisotropic ones below 0.8 K.

In the following, we discuss the temperature dependence of $1/T_1$ below $T_{\rm f}$ on the basis of a conventional two-magnon process in a triangular Heisenberg antiferromagnet. $1/T_1$ determined by the two-magnon process was discussed by Moriya,\cite{Moriya} and was extended to the triangular lattice by Maegawa.\cite{MaegawaPRB95} $1/T_1$ by this process is expressed as 
\begin{eqnarray*}
\frac{1}{T_1} &=&\frac{\pi}{2}\gamma_e\gamma_n\hbar\sum_{i,j}G_{ij}\int_{\omega_0}^{\omega _{\rm m}}\left\{1+\left(\frac{\omega_ {\rm m}}{\omega}\right)^2\right\} \\
 & & \times \frac{e^{\hbar \omega /k_{\rm B}T}}{(e^{\hbar \omega /k_{\rm B}T}-1)^2}N(\omega)^2d\omega,
\end{eqnarray*}
where $\gamma_e$ is the electronic gyromagnetic ratio, $G_{i,j}$ is a geometrical factor. $\omega _{\rm m}$ is the maximum frequency of the spin wave, $\omega_{0}$ is the spin-anisotropy energy related to the internal field at the Ni sites, and $N(\omega)$ is the state density of magnons. If we assume the long-wave length approximation, $N(\omega)$ in the 3-D and 2-D spin waves is 
\[
N(\omega)=
\begin{cases}
\frac{3\omega\sqrt{\omega^2-\omega_0^2}}{(\omega_{\rm m}^2-\omega_0^2)^{3/2}} & \mbox{3-dimensional}\\
\frac{\omega}{2\pi\omega_{\rm ex}^2~a^2} & \mbox{2-dimensional,}
\end{cases}
\]
where $\omega_{\rm ex}$ is the exchange frequency between the Ni spins.
Using $N(\omega)$, the temperature dependence of $1/T_1$ determined by the 3-D and 2-D spin waves is given by
\begin{eqnarray*}
\left(\frac{1}{T_1}\right)_{\rm 3D} & \propto & T^5\int_{T_0/T}^{T_{\rm m}/T}\left\{x^2-\left(\frac{T_ {0}}{T}\right)^2\right\}\\
 & & \left\{x^2+\left(\frac{T_ {\rm m}}{T}\right)^2\right\}\frac{e^{x}}{(e^{x}-1)^2}dx
\end{eqnarray*}
and
\[
\left(\frac{1}{T_1}\right)_{\rm 2D} \propto T^3\int_{T_0/T}^{T_{\rm m}/T}\left\{x^2+\left(\frac{T_ {\rm m}}{T}\right)^2\right\}\frac{e^{x}}{(e^{x}-1)^2}dx
\]
respectively.
Here, $T_{\rm m}=\hbar\omega_{\rm m}/k_{\rm B}$ and $T_{0}=\hbar\omega_0/k_{\rm B}$. The temperature dependences of $(1/T_1)_{\rm 3D}$ and $(1/T_1)_{\rm 2D}$ are calculated using $T_m \sim |\theta_{\rm W}| = $ 80 K, and $T_0 = 0.4$ K, which are shown in Fig.~14 by dotted and solid curves, respectively.
 
\begin{figure}[htbp]
 \begin{center}
 \includegraphics[clip=,width=0.9\columnwidth]{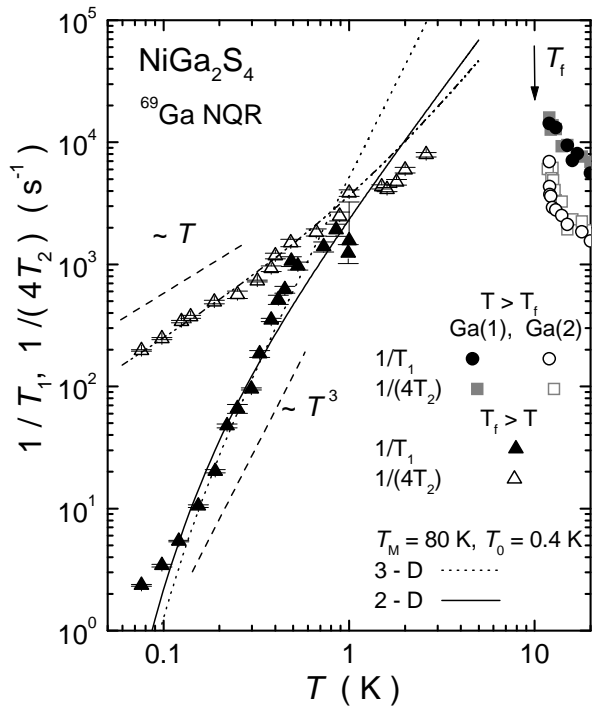}
 \caption{Temperature dependence of $1/T_1$ calculated by the 3-D and 2-D spin-wave models are compared with the experimental data (see text). }
 \end{center}
\end{figure}
$1/T_1$ data in the temperature range below 1 K is roughly reproduced by both calculations using the gap magnitude of $T_0$ = 0.4 K, however a deviation between the calculation and $1/T_1$ data is observed below 0.15 K. The deviation is considered to be due to the distribution of $T_0$, which is suggested from the inhomogeneous internal field as discussed in Sec. IV B. It seems that the 2-D model is more suitable for interpreting the whole temperature dependence below $T_{\rm f}$, because the deviation between the data and the 3-D calculation above 1 K is more significant than that in the 2-D one. $1/T_1$ from the 3-D model is approximately a $T^3$ dependence above 1 K, which is incompatible with the experimental result. The 2-D spin-wave model seems to be consistent with the $T^2$ dependence observed in specific-heat experiments above 0.35 K,\cite{NakatsujiScience} which suggests the existence of linearly dispersive modes. We point out that the 2-D spin-wave model also give the $\omega$-linear density of states. 

The temperature dependence of $1/(4T_2)$ is consistently interpreted by the dominant $T$-linear $G_z(0)$ and the small $T^3$ dependence from the $G_{\perp}(\omega)$ contribution discussed above. $1/(4T_2)$ below 1 K is approximately expressed by 
\[
\frac{1}{4T_2} = 2.5 \times 10^3 T \hspace{0.3cm} [\rm K^{-1} s^{-1}] + \frac{1}{2}G_{\perp}(\omega),
\]
which is shown by a dotted curve in Fig.~14.

In the Heisenberg triangular magnet with the 120$^{\circ}$ spin structure, there are three kinds of low-energy excitation modes, viz. swinging fluctuations in the $xy$ plane, rotational fluctuations around one of the moment direction, and vertical deviation from the $xy$ plane (see Fig.~4 in Ref.\onlinecite{MaegawaPRB95}). It is considered that the lowest excitation mode is the swinging fluctuation, because the fluctuation conserves energy.\cite{MaegawaPRB95}
Taking into account that NiGa$_2$S$_4$ has an incommensurate short-range correlation with $\mbox{\boldmath$q$}$ close to (1/6, 1/6, 0), which is a 60$^{\circ}$ spin structure,\cite{NakatsujiScience} it is reasonable to consider that NiGa$_2$S$_4$ has magnetic fluctuations similar to those in the triangular compounds. It should be noted that the out-of plane dynamics is homogeneous although the in-plane dynamics is inhomogeneously distributed. It is considered that the out-of-plane dynamics is related to the lowest excitation, which is the swinging fluctuation in the $xy$ plane, because the swinging fluctuation gives rise to $\langle S_z(t)S_z(0)\rangle$. However, it is difficult to identify what kinds of magnetic fluctuations are associated with the in-plane and out-of-plane spin dynamics at the Ga site. To determine the origin of the temperature dependence in $G_{\perp}(\omega)$ and $G_{z}(0)$, a theoretical study of the magnetic fluctuations with a 60$^{\circ}$ spin structure is desired.

One of the characteristic features of the $1/T_1$ results on NiGa$_2$S$_4$ is that $1/T_1$ retains a large value down to 1 K, i.e., far below $T_{\rm f}$, which is also indicated by the $\mu$SR relaxation rate in Fig.~10. The large value of $1/T_1$ is related to the fact that the NQR spectrum affected by the inhomogeneous static field cannot be observed between 2 K and $T_f$. In an ordinary magnetic transition and spin glass ordering, $1/T_1$ shows a divergence at the magnetic ordering temperature $T_{\rm M}$ and decreases abruptly below $T_{\rm M}$, and then the internal field emerges. Contrastingly, in NiGa$_2$S$_4$, $1/T_1$ is too short to be measured by the Ga-NQR measurements down to 2 K ($\sim T_{\rm f} / 5$). This indicates that the magnetic fluctuations are not quenched abruptly but gradually decrease or stay constant down to 2 K.

Another characteristic feature of the $1/T_1$ behavior is that the enhancement of $1/T_1$ does not start just above $T_{\rm f}$ but at 80 K, which is far above $T_{\rm f}$. Such behavior of $1/T_1$ suggests that the release of the magnetic entropy at $T_{\rm f}$ is small. It is considered that the gradual decrease of the magnetic entropy is the reason why the anomaly of the specific heat at $T_{\rm f}$ is so broad that the anomaly seems to be different from ordinary magnetic ordering. We point out that the small change of the entropy at the magnetic anomaly is one of the characteristic features of frustrated magnetism, because a similar broad maximum in the specific-heat results is often observed in not only in triangular compounds\cite{OlariuPRL06,AlexanderCondmat} but also in kagom\'e compounds.\cite{RamirezPRL90,MatsunoPRL03}

Furthermore, there remains a question to be understood when we interpret the temperature dependence of $1/T_1$ on the basis of the spin-wave model. As reported from the neutron experiment, the magnetic correlation length in NiGa$_2$S$_4$ is short with only 6 lattice constants even at 1.5 K, which makes NiGa$_2$S$_4$ resemble more conventional spin glasses than the long-range ordered 2-D AF magnet. The presence of the spin wave seems to be unrealistic in such a short-range ordered compound. In addition, the spin-wave model is considered to have a significant field dependence of $1/T_1$, which seems to be contradict with the field insensitive $C(T)$.\cite{NakatsujiScience} 

However, we point out that the similar situation was observed in SCGO,\cite{RamirezPRL90} in which the magnetic correlation length is only two times of inter-Cr spacing\cite{BroholmPRL90} but specific heat $C(T)$ varies as $T^2$ at low temperatures.\cite{RamirezPRL90} In addition, the similar slowing down of the Cr spin fluctuations when approaching to the susceptibility-cusp temperature $T_g \sim 3.5$ K was reported from the $\mu$SR and Ga-NMR measurements.\cite{UemuraSCGOPRL94,MendelsPRL00} It is considered that the linear dispersion indicated from the specific-heat behavior does not imply the presence of the long correlation length. We suggest the possibility of a cluster glass, which might be consistent with all existing results except for the field insensitive behavior of $C(T)$. The spin-wave excitations arise inside the cluster and the clusters freeze out independently. It is considered that the size of the cluster might be approximately 6 lattice constants. The persisting spin fluctuations below $T_{\rm f}$ might be associated with the spins at the cluster boundaries. To examine this scenario, ac susceptibility measurements and theoretical studies of the field dependence of $N(\omega)$ in the frustrated systems are highly desired. 

Quite recently, Tsunetsugu and Arikawa pointed out that spin nematic order can explain the experimental facts of absence of the magnetic long-range order, a power-law behavior $C \sim T^2$, and the incommensurate wave vector $\mbox{\boldmath$q$} \sim$ (1/6, 1/6, 0).\cite{TsunetsuguJPSJ06} The order parameters of the spin nematic order are not related to ordinary static spin dipole moments, but related to anisotropy of spin fluctuations. The present results of $1/T_1 \sim T^3$ and the anisotropic spin fluctuations below $T_{\rm f}$ seems to be consistent with the spin nematic order. However, it is a crucial point whether the spin freezing behavior accompanied by the internal magnetic fields below 2 K, observed by the NMR/NQR and $\mu$SR measurements, is consistent with their scenario.    

On the other hand, Kawamura and Yamamoto studied the ordering of the classical Heisenberg antiferromagnet on the triangular structure with bilinear and biquadratic interactions.\cite{KawamuraJPSJ07} They suggested that a topological phase transition at a finite temperature driven by topological vortices although the spin correlation length remains finite even below the transition point. They pointed out that the magnetic anomaly at $T_{\rm f}$ in NiGa$_2$S$_4$ might originate from a vortex-induced topological transition related to $Z_2$ vortices.\cite{KawamuraJPSJ07} In their vortex scenario, the magnetic-correlation time does not truly diverge at $T_{\rm f}$, but only grows sharply at $T_{\rm f}$ exceeding the experimental time scale, and remains short over wide temperature range below $T_{\rm f}$. Correspondingly, magnetic-correlation length $\xi$ is short and remains finite in this temperature range, and the onset of the magnetic long-range order, which is characterized by the exponential divergence of $\xi$, is observed only at still lower temperatures.\cite{KawamuraJPSJ07} 
The magnetic properties revealed with the present NQR, $\mu$SR and the neutron experiments above 2 K seem to be consistent with the above vortex transition. In addition, if the temperature for long-range order in the vortex scenario is considered to be $\sim$ 2 K, the observation of spin-wave-like excitations similar to those in a long-range ordered 2-D AF magnet can be understood. To examine this vortex scenario, the neutron experiments below 2 K are highly desired.

\section{Conclusion}

We have performed $^{69,71}$Ga-NMR/NQR and $\mu$SR measurements on triangular antiferromagnet NiGa$_2$S$_4$.
NMR and NQR spectra above $T_{\rm f}$ indicate the existence of two Ga sites with different local symmetries, although there exists only one crystallographic site in perfect NiGa$_2$S$_4$.
The intensity ratio of two NQR peaks is approximately 4 : 1 and the linewidth of the intense peak is narrower than the other.
At present, the origin of the two NQR peaks is not fully understood, but it is speculated that a tiny amount of sulfur disorder and/or planar defects such as stacking faults might induce the different Ga site.

1/$T_1$ and $1/T_2$ were measured at both Ga sites and divergent behavior of $1/T_1$ and $1/T_2$ was observed at $T_{\rm f}$ at both Ga sites.
It was found that $1/T_1$ and $1/T_2$ show the same temperature dependence above $T_{\rm f}$, indicative of isotropic spin dynamics in the rapid-motion limit. The observed isotropic spin dynamics is considered to be characteristic of a Heisenberg spin system.
With decreasing temperature, magnetic correlations start to develop below $|\theta_{\rm W}| \sim 80$ K, which is far above $T_{\rm f}$. Spin fluctuations continue to slow down below $|\theta_{\rm W}|$ and become nearly static below $T_{\rm f} = 10$ K, where the specific heat $C(T)$ shows a broad maximum. The remarkable divergence of $1/T_1$ and $1/T_2$, which is observed over wide temperature range between $\theta_{\rm W}$ and $T_{\rm f}$, is characteristic of frustrated systems, because geometrical frustration suppresses the magnetic anomaly down to low temperatures. 

However, the wide temperature region between $T_{\rm f}$ and 2 K, where the NQR signal was not observed, suggests that the Ni spins do not freeze immediately below $T_{\rm f}$, but keep fluctuating down to 2 K with the MHz frequency range. Below 0.5 K, all components of $1/T_1$ follow a $T^3$ behavior. 
Below $T_{\rm f}$, we found a broad spectrum and an inhomogeneous distribution of $T_1$ in NMR and NQR measurements.
These results indicate the freezing of magnetic moments and the emergence of inhomogeneous static magnetism below 2 K. These are also suggested from the $\mu$SR experiments.
The relaxation rates decrease below $T_{\rm f}$, and $1/T_1$ follows a $T^3$ dependence below 0.8 K, and the overall temperature dependence of $1/T_1$ is roughly interpreted by the 2-D spin-wave model.
In addition, on the basis of the observed difference in the temperature dependence of $1/T_1$ and $1/T_2$, the spin dynamics is interpreted to become anisotropic below 0.8 K. These experimental results strongly suggest that short-range magnetic order with incommensurability and/or inhomogeneous static moments is realized below $T_{\rm f}$, which is consistent with the neutron experiments. We suggest that the $\omega$-linear dispersion implied from the specific-heat measurements is understood as a consequence of the inhomogeneous magnetic fields due to the inhomogeneous short-range magnetic order. The spin dynamics in the short-range ordered state, which is considered to be field insensitive, and the magnetic correlations developing below 80 K are interesting issues in frustrated magnetism to be further studied from theoretical and experimental points of view.   

\begin{acknowledgments}
We thank  S.~Maegawa, N.~B$\rm {\ddot{u}}$ttgen, M. Takigawa, S.~Fujimoto and H.~Kawamura for valuable discussions.
This work was partially supported by CREST of the Japan Science and Technology Agency (JST) and the 21 COE program on ``Center for Diversity and Universality in Physics'' from MEXT of Japan, and by Grants-in-Aid for Scientific Research from the Japan Society for the Promotion of Science (JSPS)(No.16340111 and 18340102), MEXT(No.16076209), and US National Science Foundation (No. 0422674).
\end{acknowledgments}


\end{document}